\documentclass[twocolumn,showpacs,pra,10pt]{revtex4-1}
\usepackage[english]{babel}
\usepackage{amsmath,graphicx,amssymb} 
\usepackage{times}
\usepackage{epsfig} 
\usepackage{epstopdf} 
\usepackage{color}

\begin{document} 
\title{Effect of quantum nuclear motion on  hydrogen bonding
}

\author{Ross H. McKenzie}
\email{email: r.mckenzie@uq.edu.au}
\homepage{URL: condensedconcepts.blogspot.com}
\author{Christiaan Bekker} 
\affiliation{School of Mathematics and Physics, University of Queensland,
  Brisbane 4072, Australia} 

\author{Bijyalaxmi Athokpam and Sai G. Ramesh}
\affiliation{Department of Inorganic and Physical Chemistry,
Indian Institute of Science, Bangalore 560 012, India}

\date{\today}
                   
\begin{abstract}

This work considers how the properties of hydrogen bonded complexes,
X-H$\cdots$Y, are modified by the quantum motion of the shared proton. Using a
simple two-diabatic state model Hamiltonian, the analysis of the symmetric case,
where the donor (X) and acceptor (Y) have the same proton affinity, is carried
out. For quantitative comparisons, a parametrization specific to the
O-H$\cdots$O complexes is used. The vibrational energy levels of the
one-dimensional ground state adiabatic potential of the model are used to make
quantitative comparisons with a vast body of condensed phase data,
spanning a donor-acceptor separation ($R$) range of about $2.4-3.0$ {\AA}, i.e.,
from strong to weak hydrogen bonds. The position of the proton (which determines the
X-H bond length) and its longitudinal
vibrational frequency, along with the isotope effects in both are described quantitatively. An
analysis of the secondary geometric isotope effect, using a simple extension of
the two-state model, yields an improved agreement of the predicted 
variation with $R$ of frequency isotope effects. 
The role of bending modes is also considered: their quantum effects compete with those of the
stretching mode for weak to moderate H-bond strengths.
In spite of the economy in the parametrization of the model used, it offers key insights into the
defining features of H-bonds, and semi-quantitatively captures several trends.

\end{abstract}

\maketitle 

\section{Introduction}

In most chemical systems, nuclear quantum zero-point motion and tunneling do
not play a significant role. Most of chemistry can be understood in terms of
semi-classical motion of nuclei on potential energy surfaces.  In contrast, the
quantum dynamics of  protons involved in hydrogen bonds plays an important role
in liquid water \cite{ChenPRL03,MorronePRL08,CeriottiPNAS13}, ice
\cite{BenoitNature,Pamuk}, transport of protons and hydroxide ions in water
\cite{Tuckerman02}, surface melting of ice \cite{Paesani08}, the bond
orientation of water and isotopic fractionation
 at the liquid-vapour interface \cite{Nagata}, isotopic
fractionation in water condensation \cite{Markland}, proton transport in
water-filled carbon nanotubes \cite{Chen}, hydrogen chloride hydrates
\cite{Hassanali}, proton sponges \cite{Bienko,Horbatenko}, water-hydroxyl
overlayers on metal surfaces \cite{LiPRL10}, and in some proton transfer
reactions in enzymes \cite{BothmaNJP10}.  Experimentally, 
the magnitude of these nuclear quantum effects
are reflected in isotope effects, where hydrogen is replaced with deuterium.

The quantum effects are largest for medium to strong symmetric bonds
where the proton donor (X) and acceptor (Y) are identical (i.e., have the same
proton affinity) and are separated by distances ($R$) of about $2.4-2.5$ \AA.
In a  recent review about the solvation of  protons,
Reed noted the importance
of this parameter regime: ``In contrast to the typical asymmetric H-bond found in proteins (N–H$\cdots$O) or ice (O–H$\cdots$O), the short, strong, low-barrier (SSLB) H-bonds found in proton disolvates, such as H(OEt$_2)_2^+$ and H$_5$O$_2^+$, deserve much wider recognition'' \cite{Reed}.

The approach of this paper is to consider a simple, physically insightful model,
and to probe the extent to which the quantum treatment of the one-dimensional
proton motion afforded by it modifies properties of hydrogen bonded complexes.
While the model is general, we specifically target O-H$\cdots$O bonds for quantitative comparisons. 
We compare the predictions of the model to
a large body of experimental data, where the O$\cdots$O distance spans
a range from about 2.4 \AA~(strong bonding) to 3.0 \AA~(weak bonding).

Many other works using multi-dimensional potential energy surfaces, parametrised
by {\it ab initio} calculations for specific molecular complexes, have been
carried out earlier. However, such studies are computationally rather demanding.
The present work is intended to complement such studies: we attempt to demonstrate
that much of the crucial physics can be described by a one-dimensional quantum
treatment alone. But, we also show this treatment cannot describe the secondary
geometric isotope effect for weak to moderate bonds; inclusion of bending vibrations is 
necessary.

The outline of the paper is as follows.  In Section \ref{PES}, we describe a
simple potential energy surface based on a two-diabatic-state model,  considered in
Ref.~\onlinecite{McKenzieCPL} recently. This potential has the key property that it undergoes
qualitative changes as $R$ varies between 2.4 {\AA} and 2.6 \AA. We focus on its
one-dimensional slices along the linear proton path between the donor and the
acceptor. 
 Vibrational eigenstates obtained for these
slices for a large range of X-Y separations (Section \ref{sec-vib}) are used to
analyse various properties of the H-bond and compare the results with
experiment.  Section \ref{sec-bondlengths} presents the
modification of the X-H bond lengths.  Section \ref{sec-freq} considers the
correlation between the X-H stretch frequency and the donor-acceptor distance,
showing the importance of anharmonic effects.
Section \ref{sec-isotope} discusses geometrical and vibrational frequency
isotope effects; they are largest when the zero-point energy is comparable to
the height of the potential barrier for proton transfer.  We show that for strong to moderate bonds,
the secondary geometric isotope effect is dominated by the $R$ dependence of
the zero-point energy associated with the X-H stretch mode.
Simple model potentials provide
some insight into the trends in the isotope effects that are observed as $R$ is
varied.
Section \ref{sec-compete} discusses how
description of the secondary geometric isotope effects
for weak bonds requires inclusion
of the competing quantum effects associated with the zero-point motion
of the  bending vibrational modes.

\begin{figure}[htb] 
\centering 
\includegraphics[width=84mm]{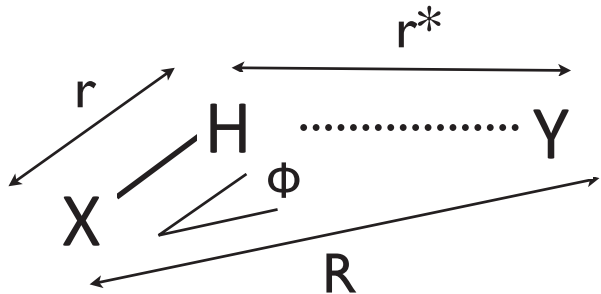}
\caption{
Definition of geometric variables for a hydrogen bond between
a donor (X) and an acceptor (Y).
This paper is concerned with the quantum motion of the proton
H relative to X and Y. The focus in on the case of
linear bonds where $\phi=0$ and $r^*=R-r$.
 The quantum effects are largest when the 
donor-acceptor distance $R$ is about $2.4-2.5$ \AA.
}
\label{fig1}
\end{figure}

\section{
A simple model for ground state potential energy surfaces
}
\label{PES}

This is based on recent work by one of us \cite{McKenzieCPL}.
We briefly review the underlying physics and chemistry
behind the simple effective Hamiltonian which produces the 
potential energy surfaces that we use to describe the nuclear motion.

\subsection{Reduced Hilbert space for the effective Hamiltonian}

Diabatic states \cite{PacherJCP88}, including valence bond states, are a
powerful tool for developing chemical concepts \cite{Shaik}.  It has
been proposed in many earlier works that hydrogen bonding and hydrogen transfer reactions can be
described by an Empirical Valence Bond (EVB)
model \cite{Horiuti,CoulsonAF54,Warshel,BorgisJMS97,SagnellaJCP98,FlorianJPCA02}
involving valence bond states.  In the  present case, the reduced Hilbert space has a basis
consisting of two diabatic states that can be denoted as 
$\left|\text{X-H}^+,\text{Y}\right\rangle$ and 
$\left|\text{X},\text{H-Y}^+\right\rangle$.  The latter
represents a product state of the electronic states of an isolated X molecule and of an isolated Y-H$^+$ molecule.
 The difference between the two diabatic states is
transfer of a proton from the donor to the acceptor.  
Note that the positive charges in this notation are nominal,
only indicating the presence of the transferring proton on X or Y. The total
charge on each of X-H and Y-H would, of course, depend on the charges of X and Y
themselves. The X-H and H-Y bonds have both covalent and ionic
components, the relative weight of which depends on the distances $r$ and $r^*$, respectively. To illustrate these diabatic states we
consider three specific examples.

\begin{enumerate}
\item For the Zundel cation, (H$_5$O$_2$)$^+$, a proton is transferred between
two water molecules, X$\,=\,$Y$\,=\,$H$_2$O.  The two diabatic states are 
$|$H$_3$O$^+$, H$_2$O$\rangle$ and 
$|$H$_2$O, H$_3$O$^+\rangle$ which are degenerate.

\item For the (H$_3$O$_2$)$^-$ ion, a proton is transferred between
two hydroxide anions: X$=\,$Y$=\,$OH$^-$.
The two diabatic states are 
$|$H$_2$O, OH$^-$$\rangle$ and 
$|$OH$^-$, H$_2$O$\rangle$, 
which are degenerate.

\item Hydrogen bonding between two water molecules, can viewed
in terms of  proton
 transfer between a water molecule and a hydroxide anion: 
X$=$ OH$^-$ 
and
Y$\,=\,$H$_2$O, 
and so this is an asymmetric case.
The two diabatic states are 
$|$H$_2$O, H$_2$O$\rangle$ and 
$|$OH$^-$, H$_3$O$^+\rangle$, which are non-degenerate.
A very crude estimate of 
the energy difference between these two states,
neglecting significant solvation effects present
 in aqueous solution,
is the free energy difference 
21 kcal/mol corresponding to an equilibrium constant of 
$10^{-14}$. 
\end{enumerate}

In this paper, we focus solely on the the symmetric case where
the donor and acceptor have the same proton affinity.

\subsection{Effective Hamiltonian}

The  Hamiltonian for the two diabatic states
has matrix elements that depend on
the X-H bond length $r$, the donor-acceptor separation $R$,
and the angle $\phi$, which describes the deviation from linearity 
(compare Figure \ref{fig1}).
It was recently shown that
one can obtain both a qualitative and semi-quantitative
description of hydrogen bonding using
a simple and physically transparent parametrisation
of these matrix elements \cite{McKenzieCPL}.
This approach unifies H-bonding involving different atoms and
weak, medium, and strong (symmetrical) H-bonds.

The Morse potential describes the energy of a single bond 
within one of the molecules in the absence of the second 
(and thus the diabatic states). 
A simple harmonic potential is not sufficient because
the O-H bond is highly anharmonic and we
will be interested in regimes where there is considerable
stretching of the bonds.
The two cases $j=$X, Y denote the
donor X-H bond and acceptor Y-H bond, respectively.
The Morse potential is
\begin{equation}
V_j(r) = D_j\left[e^{-2a_j(r-r_{0j})} - 2e^{-a_j(r-r_{0j})}\right],
\label{morse}
\end{equation}
where $D_j$ is the binding energy, $r_{0j}$ is the equilibrium bond length,
and $a_j$ is the decay constant. 
$D_X$ and $D_Y$ denote the proton affinity of the donor
and the acceptor, respectively.
For O-H bonds, approximate parameters are $D \simeq 120$ kcal/mol,
$a \simeq 2.2$ \AA$^{-1}$, $r_0 \simeq 0.96$ \AA, which correspond to an O-H
stretch harmonic frequency, $\omega$, of $\simeq 3750$ cm$^{-1}$.

We take the effective Hamiltonian describing the two interacting
diabatic states to have the form
\begin{equation}
H = \left(\begin{array}{cc} 
V_X (r)           & 
\Delta_{XY}(R,\phi) \\[2mm]
\Delta_{XY}(R,\phi)  & V_Y(r^*)             
\end{array}\right),
\label{eqn-ham}
\end{equation}
where
\begin{equation}
 r^* = \sqrt{R^2+r^2-2rR\cos\phi} 
\label{r*length}
\end{equation}
is the length of the Y-H bond
 (see Figure \ref{fig1}).
The diabatic states are coupled  via the off-diagonal matrix element
\begin{equation}
\Delta_{XY}(R,\phi  )=
\Delta_1 \cos\phi 
\frac{(R-r\cos\phi)}{r^*}
 e^{-b (R-R_1)}
\label{DeltaR}
\end{equation}
 (see Figure \ref{fig1}),
and $b$ defines the decay rate of the matrix element with
increasing $R$.
 $R_1$ is a reference distance that we take as $R_1 \equiv 2r_0+1/a
\simeq 2.37$ \AA.
This is introduced so that the constant $\Delta_1 $ 
sets an energy scale that is physically relevant.
The functional dependence on $R$ and $\phi$ can be justified from
orbital overlap integrals \cite{MullikenJCP49}
together with a valence bond theory description of 
four-electron three-orbital 
systems (see page 68 of Ref.~\onlinecite{Shaik}).
There will be some variation in
the parameters $\Delta_1$ and $b$ with the chemical
identity of the atoms (e.g. O, N, S, Se, ...) in the donor
and acceptor that are directly involved in the H-bond.  

\subsection{Parametrisation of the diabatic coupling}

Since the Morse potential parameters are those of isolated X-H and Y-H bonds,
the model has essentially two free parameters, $b$ and $\Delta_1$.
These respectively set the length and energy scales associated with the interaction 
between the two diabatic states.
That only two parameters are used here is in contrast to most multi-parameter
EVB models and empirical ground state potential energy surfaces \cite{Korth}.
For example, one version of the latter involves 11 parameters for
symmetric bonds and 27 parameters for asymmetric bonds \cite{LammersJCC07}.
A significant point of Ref.~\onlinecite{McKenzieCPL} was that just the
two  parameters, $b$ and $\Delta_1$,  are 
sufficient to obtain a semi-quantitative description of a wide range of experimental data for a chemically diverse set of complexes.
The parameter values that are used here,
$\Delta_1=0.4D \simeq$ 2 eV 
and $b=2.2$ \AA$^{-1}$ for O-H$\cdots$O systems,
were estimated from comparisons of the predictions
of the model
with experiment \cite{McKenzieCPL}.

\subsection{Potential energy surfaces}

In the adiabatic limit, 
 the electronic energy eigenvalues 
of Eq.~(\ref{eqn-ham}) for linear bonds ($\phi=0$) are the eigenvalues of the effective
Hamiltonian matrix:
\begin{eqnarray}
\epsilon_\pm(r,R) &=& 
\tfrac{1}{2}\left[V_X(r) + V_Y(R-r)\right] \nonumber \\
&&\pm 
\tfrac{1}{2}\left[(V_X(r) - V_Y(R-r))^2 + 4\Delta(R)^2
\right]^{\frac{1}{2}}.
\label{energies}
\end{eqnarray}
In this paper, we focus on the case of symmetric bonds where the parameters in
$V_X$ and $V_Y$ are identical. 

\begin{figure}[htb] 
\centering 
\includegraphics[width=80mm]{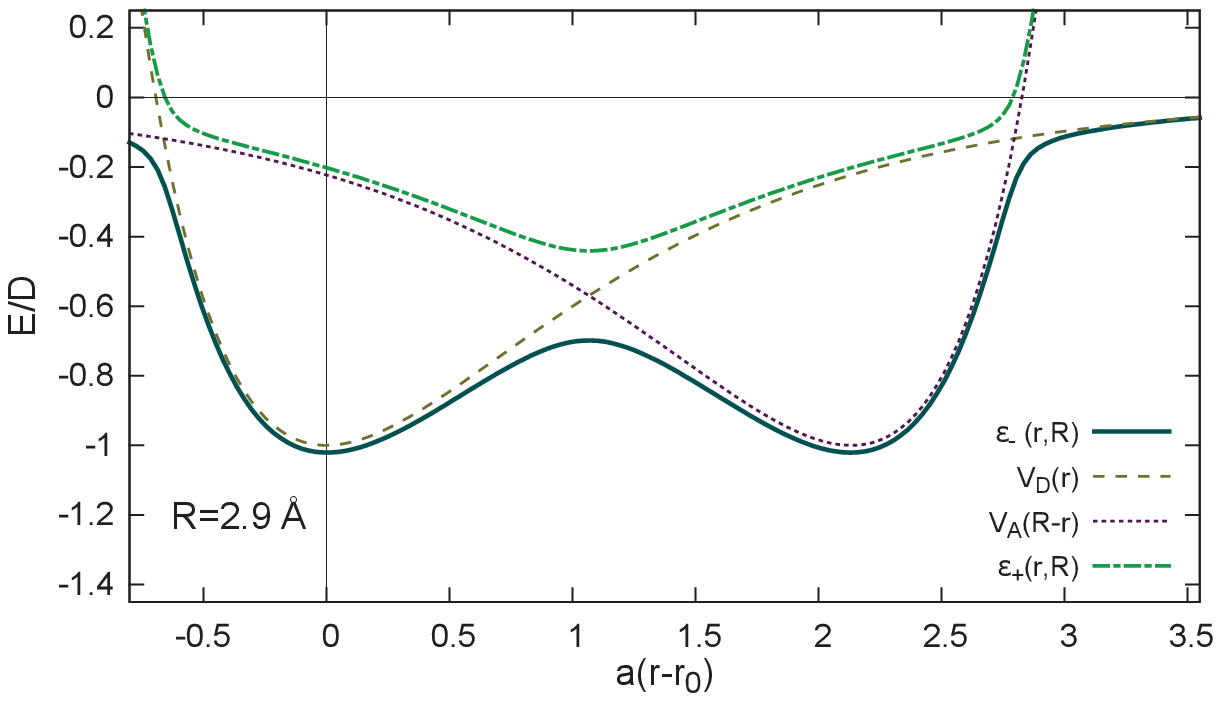}
\includegraphics[width=80mm]{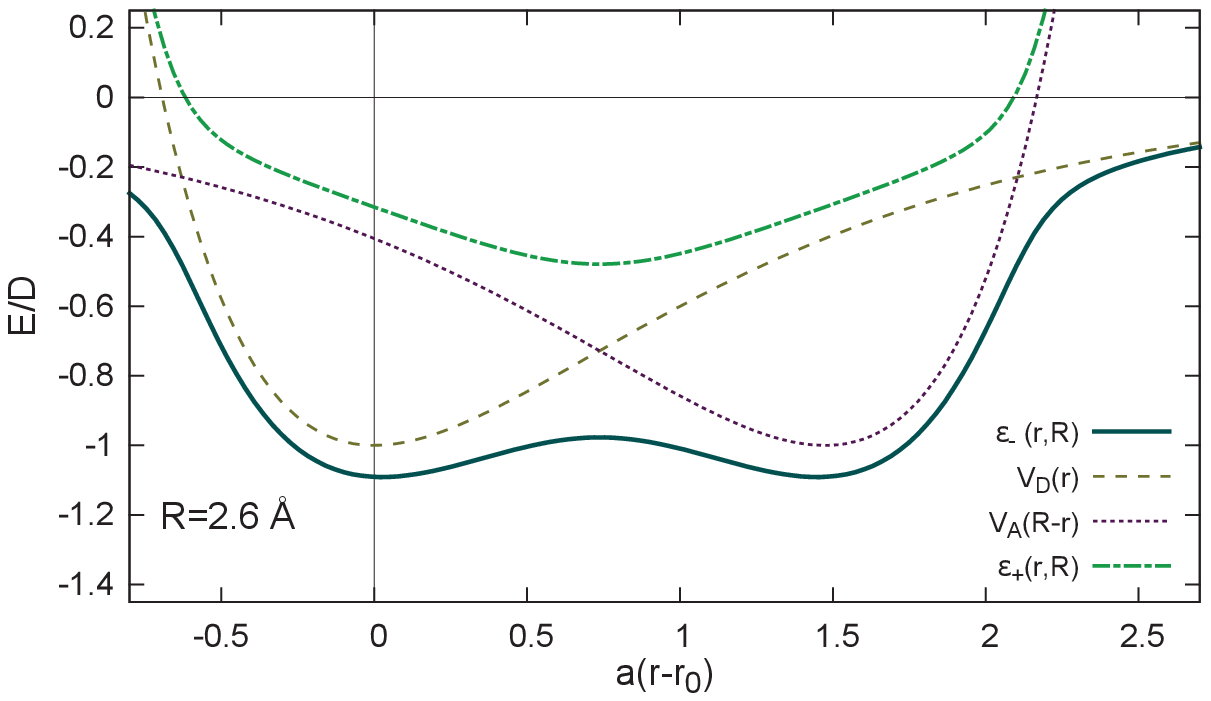}
\includegraphics[width=80mm]{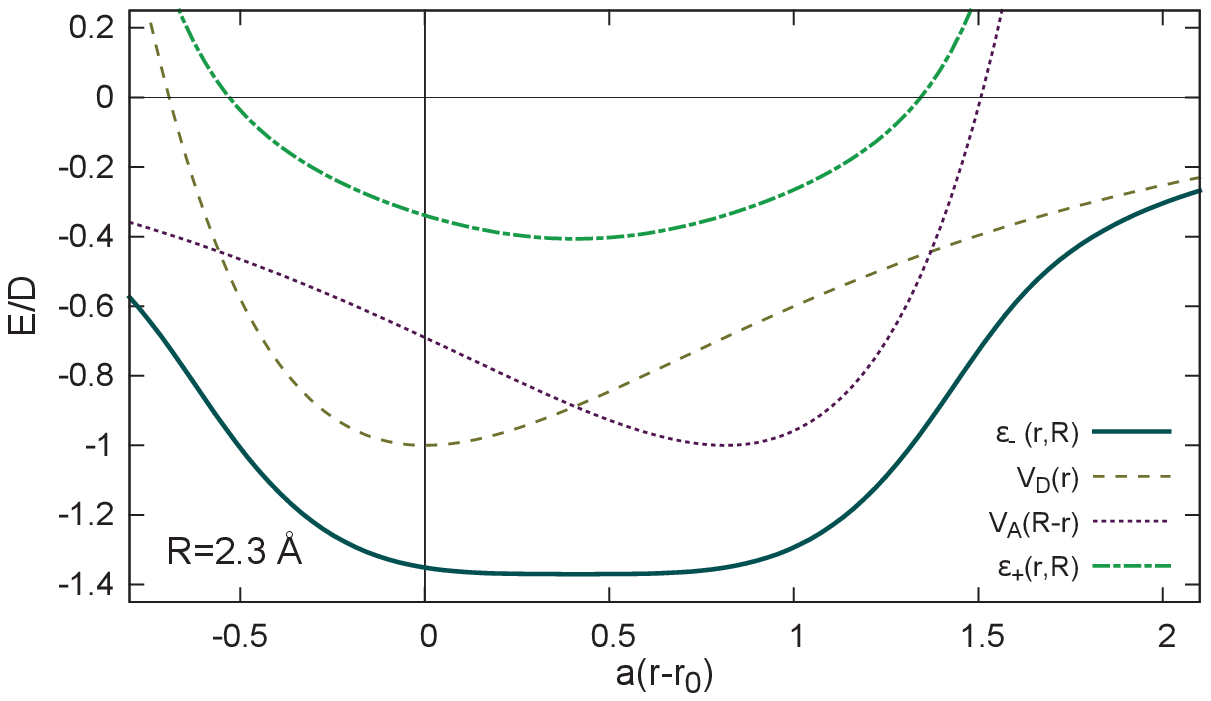}
\caption{
Potential energy curves for the diabatic and adiabatic states
of a symmetric hydrogen bonded system.
The horizontal axis is proportional to the extent of stretching 
of the X-H bond.
The vertical energy scale is $D$, the binding energy of an isolated
X-H bond.
The adiabatic curves are for an off-diagonal
coupling with parameters $\Delta_1 =0.4 D$ and $b=a$.
The diabatic curves 
are Morse potentials centred at $r=r_0$ 
(dashed lines) 
and $r^*=R-r_0$ 
(dotted lines) 
and correspond to
isolated X-H and H-Y bonds, respectively.
For parameters relevant to a O-H$\cdots$O system,
the three sets of  curves correspond (from top to bottom) 
to oxygen atom separations of
$R=2.9$, 2.6, and 2.3 {\AA}, respectively,
characteristic of weak, moderate (low barrier),
 and strong hydrogen bonds \cite{Gilli09}.
Note that the upper two panels differ from the corresponding 
figure in Ref. \cite{McKenzieCPL} due to an error in that work.
}
\label{fig2}
\end{figure}

Figure \ref{fig2} shows 
 the eigenvalues (potential energy curves)
$\epsilon_{-}(r,R)$
and
$\epsilon_+(r,R)$
as a function of $r$,  for three different fixed $R$ values.
These are three qualitatively different curves,
corresponding to weak, moderate, and strong hydrogen bonds, and
are discussed in more detail below.
[Note that Figure 2 of Reference \onlinecite{McKenzieCPL} contained
an error in the plots of the potential energy curves
and so the corrected curves are shown here.]
The surface $\epsilon_+(r,R)$ describes an
electronic excited state, and should be observable in UV
absorption experiments \cite{McKenzieCPL}. This excited state is
seen in quantum chemical calculations for the
Zundel cation \cite{Geru}.

\section{Vibrational eigenstates}
\label{sec-vib}

Under the Born-Oppenheimer approximation, the 
nuclear dynamics is determined by the adiabatic
electronic ground state potential energy,
$\epsilon_-(r,R)$.
We numerically solve the one-dimensional Schr\"odinger equation
for motion of a nucleus (proton or
deuteron) of reduced mass $M$ in this potential $\epsilon_-(r,R)$
for different fixed donor-acceptor distances $R$,
\begin{equation}
\left( -{\hbar^2 \over 2 M} { d^2 \over dr^2}  + 
\epsilon_-(r,R) \right)
\Psi_n(r)=
E_n \Psi_n(r),
\end{equation}
to find the low-lying vibrational eigenstates $\Psi_n(r)$
and energy eigenvalues $E_n$.
Isotope effects arise because the solutions depend on $M$ (see note
\cite{reduced}).
Two different numerical methods were used in order to check the results, {viz.}
the Discrete Variable Representation (DVR)\cite{Miller,Groenenboom}
with a basis of sinc-functions,
and the FINDIF program \cite{Glendening}.

\begin{figure}[!htb] 
\centering  
\includegraphics[width=73mm]{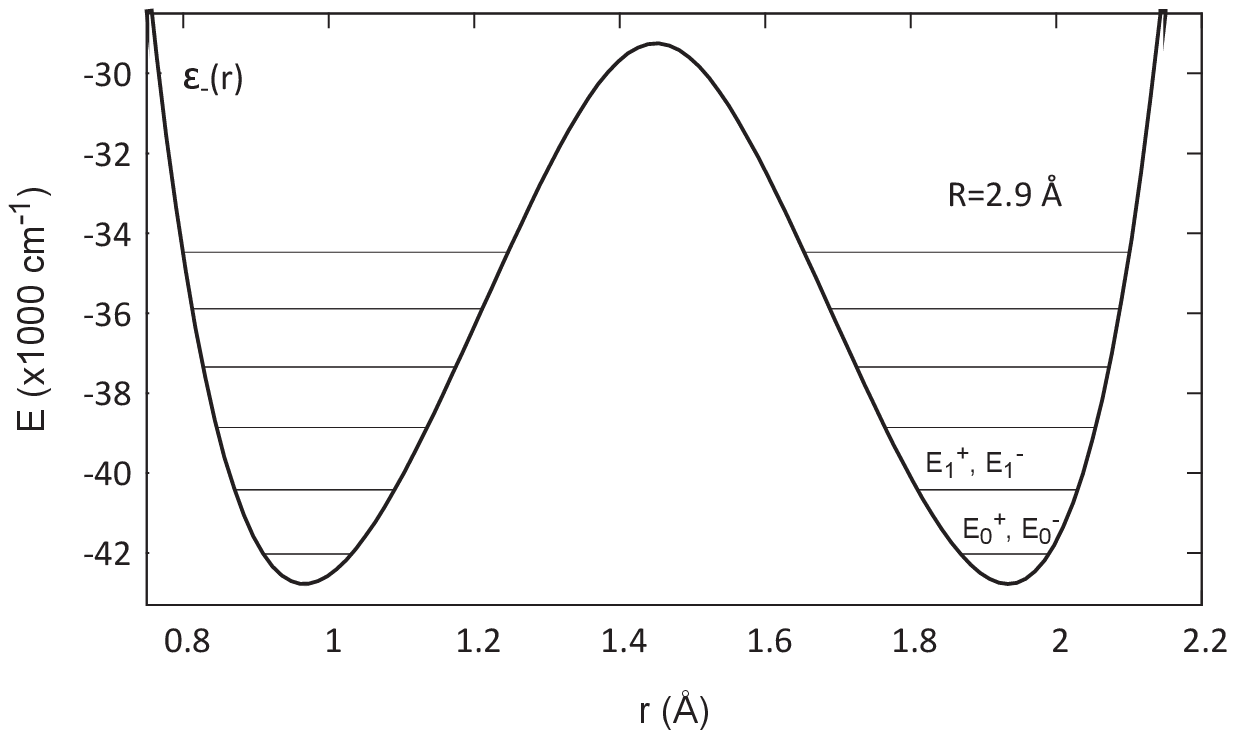}   
\includegraphics[width=72mm]{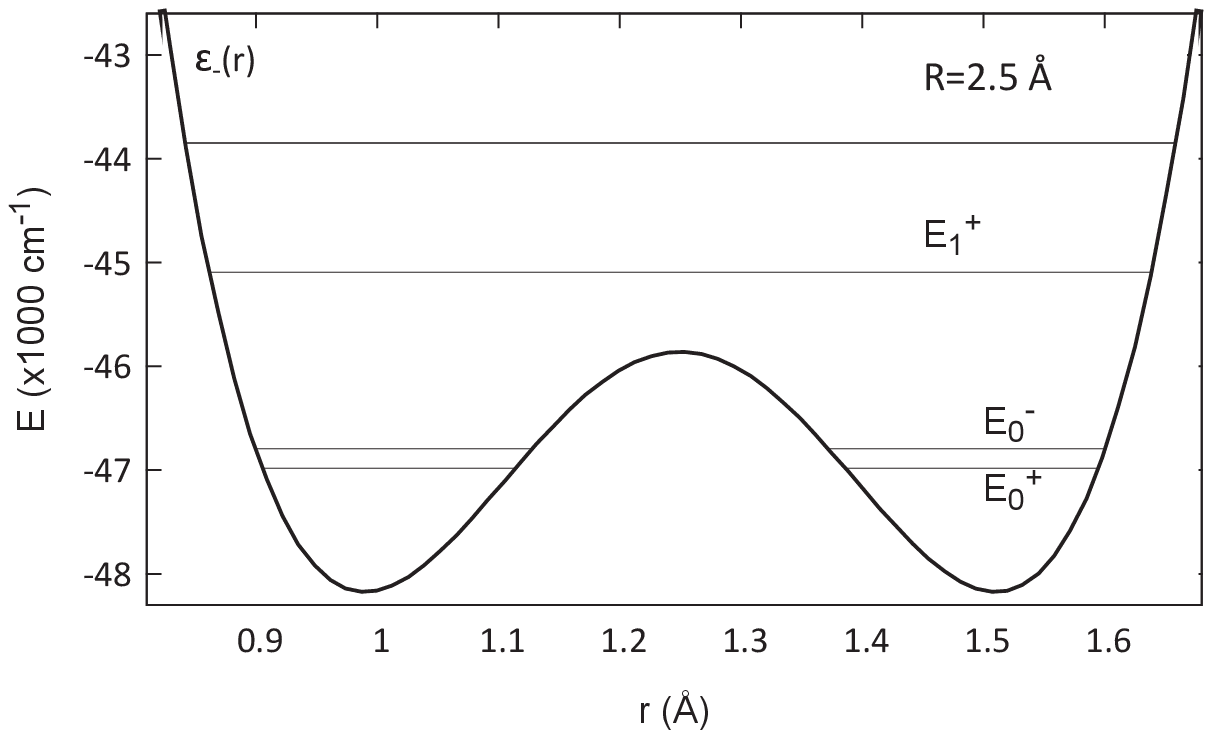}
\includegraphics[width=72mm]{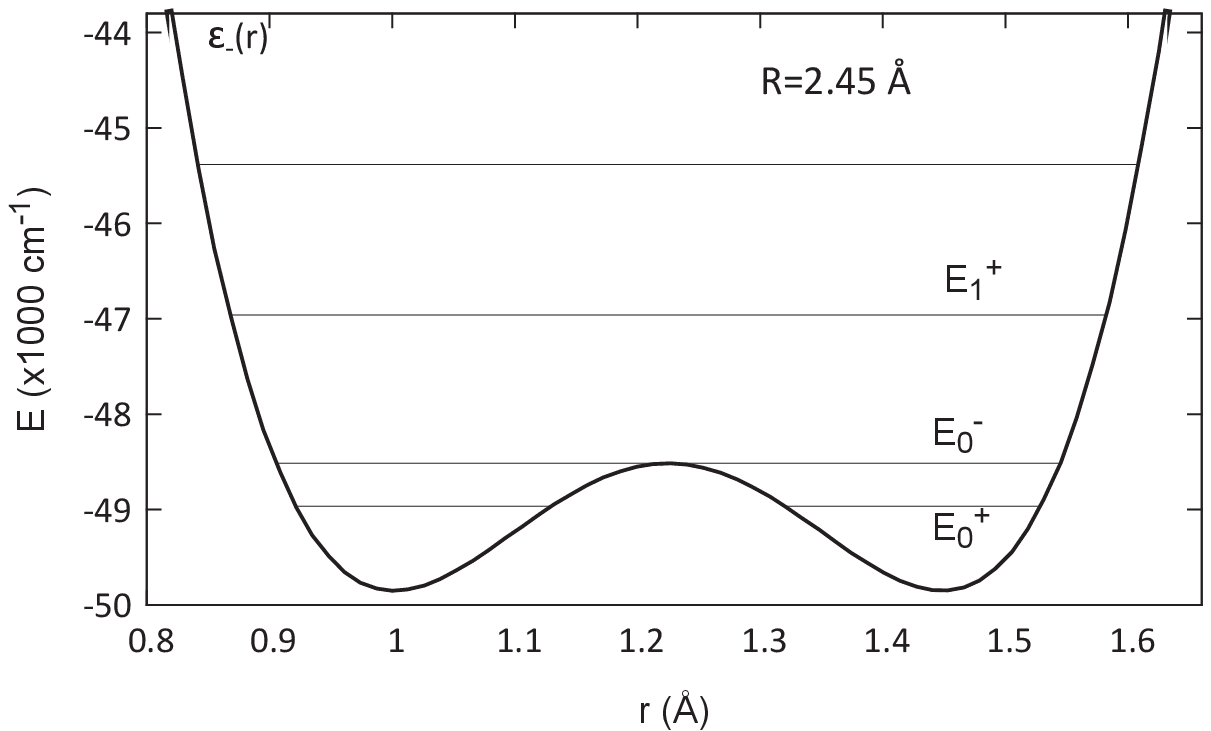}
\includegraphics[width=73mm]{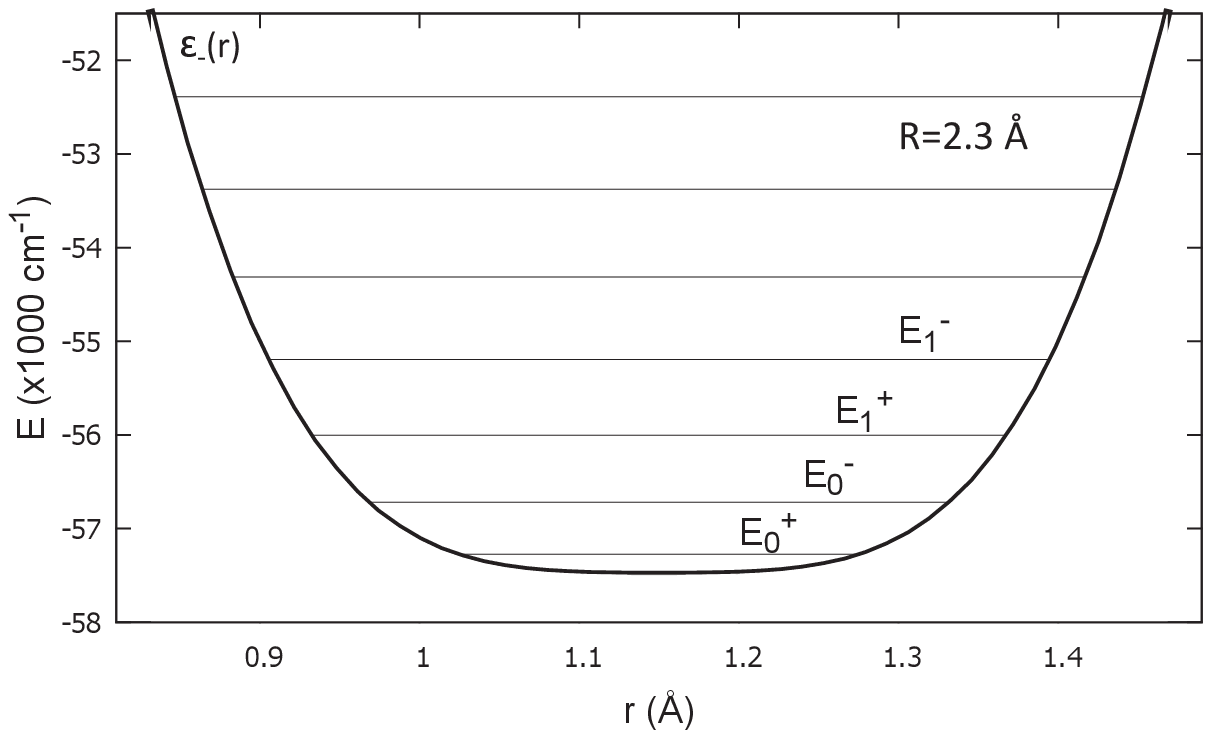}   
\caption{
Evolution of the potential energy curve and the vibrational
energy levels with decreasing donor-acceptor distance $R$,
from top to bottom.
This variation
 corresponds to changing from weak to moderate to strong symmetric bonds.
The energy levels shown are for protons.
Note that for $R > 2.6$ \AA, the
tunnel splitting between the two levels localised
on opposite sides of the potential barrier is not visible. 
In contrast, at shorter $R$, the zero-point energy becomes comparable to the barrier height and
 the tunnel splitting between the two lowest levels becomes visible.
Note that the horizontal and vertical scales of the above graphs are
slightly different from one another.
}
\label{figlevels}
\end{figure}

Figure \ref{figlevels} illustrates how the vibrational
 energy eigenvalues vary as the donor-acceptor
distance $R$ is varied.
There are three qualitatively distinct regimes:
\begin{enumerate}
\item \emph{Weak bonds} ($R >  2.6$ \AA)\\
There is a large potential barrier,
and so the tunnel splittings are a small fraction of the energy spacings.
They are not visible 
for any  of the levels on the scale of the plot shown for $R=2.9$ {\AA}
in Figure \ref{figlevels}. Nevertheless, in the gas phase small tunnel
splittings have been observed for malonaldehyde (26 cm$^{-1}$)
and tropolone (1 cm$^{-1}$)
and their derivatives \cite{Redington}.

\item \emph{Low-barrier bonds} ($R \simeq 2.4-2.6$ \AA)\\
The zero-point energy is comparable to (but less than) 
the potential barrier. 
There is a visible tunnel splitting of the two lowest levels.
The role of such bonds
 in enzyme catalysis is controversial \cite{Cleland,Warshel96,Hosur}.

\item \emph{Strong bonds} ($R \lesssim   2.4$ \AA)\\
The ground state lies above the barrier or there is no barrier \cite{note}.
All the vibrational energy levels are well-separated.
\end{enumerate}

\subsection{Proton probability density}

The relevance of the probability density to X-H bond lengths is discussed in the
next section.  As an aside, we note that the spatial probability density of the
ground state, $\rho(r)=|\Psi_0(r)|^2$, is the Fourier transform of the momentum
density $n(p)$ along the direction on of the X-H bond. A directional average of
this quantity  can be measured by deep inelastic neutron scattering
\cite{Mayers}. 
The momentum probability density has been observed
for a wide range of systems including
liquid water, ice, supercooled water, water confined
in silica nanopores \cite{Garbuio}, water at the surface of proteins
\cite{Pagnotta},
water bound to DNA \cite{Reiter10}, 
water inside carbon nanotubes \cite{Reiter06}, 
the ferroelectric KH$_2$PO$_4$ \cite{Reiter02},
hydrated proton exchange membranes \cite{Reiter12},
and a superprotonic conductor Rb$_{3}$H(SO$_{4}$)$_{2}$ \cite{Homouz}.

For all the radial distributions, $p^2 n(p)$ has a peak
for $p \sim 7$ \AA$^{-1}$.
However, with the exception of Rb$_{3}$H(SO$_{4}$)$_{2}$, liquid water, and ice, a shoulder 
or second peak is seen at larger momentum,
$p \sim 15-20$  \AA$^{-1}$.
Taking the Fourier transform leads to a real-space
ground state probability density that is bimodal, as a result
of the second peak. It can be
fitted with two Gaussians with peaks about $0.2-0.3$ {\AA} apart.
Furthermore, with knowledge of the average kinetic energy
and the probability density one can construct an effective
one-body one-dimensional potential energy for the motion
of the proton along the hydrogen bonding direction.
For the bimodal distributions the potential is a double well,
whereas for the superprotonic conductor it is narrow single well \cite{Homouz}.

These experimental results can be compared to the one-dimensional potentials
and ground state wave functions that we present here.
The comparison suggests that in 
the systems with bimodal distributions that
there is some fraction of the water molecules that are sufficiently
close that the oxygen-oxygen distance is about $2.4-2.5$ \AA. 
For reference, in bulk water this distance is about $2.8-2.9$ \AA.
However, it is possible the water molecules
could be forced closer to one another
due to the interaction of the water
with the relevant surface via bonding to the
surface or by making the water acidic or basic [producing 
H$_5$O$_2^+$ or 
H$_3$O$_2^-$ units]. Indeed, both effects occur        
for water-hydroxyl overlayers on transition metal surfaces \cite{LiPRL10}.
However, atomistic simulations of some of these specific systems
[e.g., water in silica pores \cite{Gallo}] do not seem to produce this effect.

 The probability density has been calculated for various phases of water by path
integral techniques by Morrone, Lin, and Car, using potential energy functions
from electronic structure calculations based on density functional theory
\cite{MorroneJCP}.  For water, they considered three different donor-acceptor
distances of 2.53, 2.45, and 2.31 {\AA}, corresponding to three different high
pressure phases of ice, VIII, VII, and X, respectively.  
 In Ref. \onlinecite{MorroneJSP}
these results have been
interpreted in terms of a simple empirical one-dimensional model potential.
But it was also suggested that a single proton distribution is problematic due to 
proton correlations such as those associated with the ``ice rules''.
Perhaps such effects could be treated here in a rather limited fashion
by allowing for donor-acceptor asymmetries.

\section{Bond lengths}
\label{sec-bondlengths}

\begin{figure}[htb] 
\centering  
\includegraphics[width=80mm]{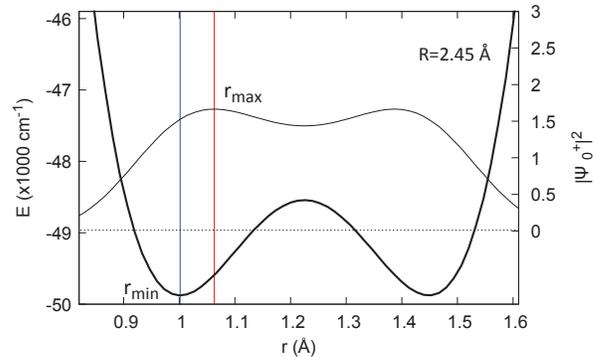}   
\caption{
Two definitions of the X-H bond length. 
One is $r_{min}$, the minimum of the ground state
adiabatic potential, a classical definition,
shown as a blue vertical line. The other, $r_{max}$, shown
as a red vertical line, is the
maximum of the ground state vibrational probability density (right-hand scale),
which accounts for the quantum vibrational zero-point motion in the anharmonic
and asymmetric (about $r_{min}$) ground state potential.  The plot is for $R =
2.45$ \AA, which falls in the moderate-to-strong  hydrogen
bond range. The dotted horizontal
line is the zero-point energy.
}
\label{figrmax}
\end{figure}

Classically, the X-H bond length is simply defined by $r_{min}$, the minimum in
the ground state potential energy.  However, if the quantum motion of the proton
is taken into account, there are ambiguities in defining the bond length that is
measured in a neutron scattering experiment.  Presumably, this bond length is
some sort of motional average associated with the ground state probability
density.  One possibility, then, is to define the bond length by $r_{max}$, the
maximum in the probability density (square of the wave function) for the proton.
If the potential energy is not symmetric about the minimum, as is the case here,
the maximum of the probability density does not correspond to the minimum of the
potential energy; this difference has been pointed out previously by Sokolov,
Vener, and Saval'ev \cite{Sokolov1}. These two different
definitions of the X-H bond length are illustrated in
Figure \ref{figrmax} for moderate-to-strong bonds.

\begin{figure}[htb] 
\centering  
\includegraphics[width=80mm]{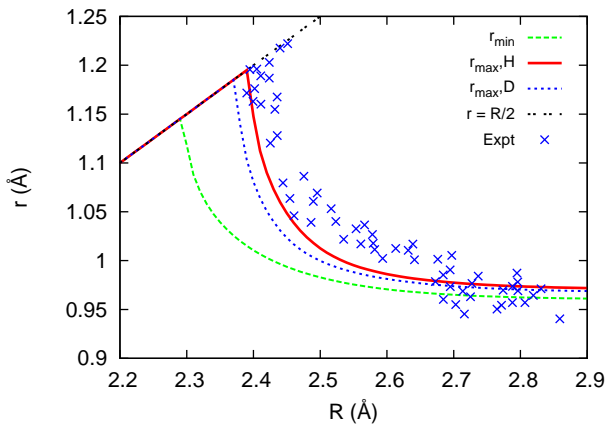}   
\caption{
Correlation between the X-H bond length $r$, defined in two ways, and the X-Y
distance $R$. 
The green dot-dashed curve, $r_{min}$, is the classical bond length (minimum of
the adiabatic ground state potential; see Figure \ref{figrmax}). The solid red and
blue dashed curves are the bond lengths are maxima of the ground state
probability distribution for the hydrogen and deuterium vibrational
wavefunctions, respectively; see Figure \ref{figrmax}.
The blue crosses are experimental data for O-H$\cdots$O bonds in a wide range of
crystal structures, and are taken from Figure 6 in 
Ref. \onlinecite{GilliJACS94}.  The
black dotted line corresponds to symmetric H-bonds ($r=R/2$) that occur when the
potential has a single minimum.
}
\label{figbondlength}
\end{figure}

Figure \ref{figbondlength} shows how quantum nuclear motion significantly shifts
$r_{max}$ (red solid line for the hydrogen and blue dashes for the deuterium)
from $r_{min}$ (green dot-dashes) as a function of the donor-acceptor distance.
The blue crosses are experimental data in Figure 6 of
Ref.~\onlinecite{GilliJACS94} for O-H$\cdots$O bonds in a wide range of
crystal structures.
A useful length scale for comparison is the zero-point amplitude of an isolated
X-H bond vibration, which is about 0.1 {\AA} for O-H bonds with ca.~3600
cm$^{-1}$ harmonic frequencies. Relative to this metric, the two bond length
definitions give distinct trends in Figure \ref{figbondlength}; 
 the $r_{max}$ curve corresponds more closely to the
measured X-H bond lengths. For the
moderate-to-strong H-bonds that occur for $R \lesssim 2.5$ {\AA}, $r_{max}$
increases more sharply because the energy barrier becomes comparable to the zero
point energy. Furthermore, there are significant primary geometric isotope
effects in the same $R$ range, i.e. the $r_{max}$ traces are significantly
different for hydrogen and deuterium.
In subsequent sections, $r_{max}$ is referred to as the X-H bond length. 

We note that similar curves to those shown in Figure 5 were produced from {\it ab initio} path integral
calculations for ice under pressure \cite{BenoitNature}.  In particular, the
transition to symmetric bonds for $R < 2.4$ {\AA} was identified with the
experimentally observed transition to ice X for pressures above 62 GPa for
H$_2$O and 72 GPa for D$_2$O \cite{Pruzan}.  Similar empirical curves including
the correction due to quantum zero-point motion have been presented for both
oxygen [O-H$\cdots$O] and nitrogen systems [N-H$\cdots$N] by Limbach and
collaborators \cite{LimbachJMS04,Limbach09}.

\section{Longitudinal vibrational frequencies}
\label{sec-freq}

There is some subtlety in using the calculated vibrational energy levels to
deduce the vibrational frequency that is actually measured in an infra-red
spectroscopy experiment. A good quantum number is the parity of the vibrational
energy level, associated with inversion symmetry about $r=R/2$ in the potentials
shown in Figure \ref{figlevels}. Each pair of tunneling-split levels have
opposite parity, and can therefore be labelled $0^+$, $0^-$, $1^+$, $1^-$, ...
(following the case of the umbrella inversion mode in ammonia
\cite{Glendening}).  The transition dipole operator has odd parity, which,
coupled with room or lower temperature Boltzmann weights of the vibrational
energy levels, suggest that the relevant transitions are $0^- \rightarrow
1^+$ and $0^+ \rightarrow 0^-$.

\begin{figure}[htb] 
\centering
 \includegraphics[width=78mm]{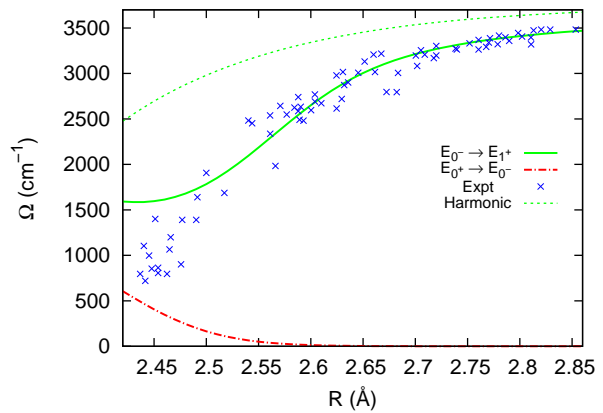}
\caption{
Softening of the X-H stretch frequency $\Omega$ (in cm$^{-1}$) with decreasing
donor-acceptor distance $R$ (in {\AA}).  The  green dashed curve is the harmonic
frequency at the $r_{min}$ of $\epsilon_-(r,R)$.
The red dot-dashed curve is the energy difference
($E_{0^-}-E_{0^+}$) 
between the two lowest lying energy levels (tunnel splitting of the
ground state).  The green solid curve is the energy difference 
($E_{1^+}-E_{0^-}$) 
between the first and second excited state energy levels. The blue crosses are experimental data for a wide range of
complexes, and are taken from Figure 4 in Ref. \onlinecite{SteinerACIE}.
}
\label{figfreq}
\end{figure}

Figure \ref{figfreq} compares the frequencies of both transitions from our
calculations with
experimental data.  As the donor-acceptor distance decreases, there is a
significant softening in the experimental X-H stretch frequency (blue crosses)
\cite{Novak,Gilli09,SteinerACIE,Libowitzky,Sikka,caveat},
 a trend that is largely traced by
the $0^- \rightarrow 1^+$ energy gap (green solid line).  This softening has
been proposed as a measure of the strength of an H-bond \cite{LiPNAS11}. The
harmonic limit, i.e. the frequency obtained from the curvature with respect to
$r$ at the bottom of the potential $\epsilon_{-}(r,R)$, is larger in value, and
an increasingly poor estimator of transition frequencies with increasing
anharmonicity (decreasing $R$) \cite{harmonic}.

The $0^+ \rightarrow 0^-$ (red dot dashed line) 
transition is of relevance only at $R
\lesssim 2.5$ {\AA}. For $R \gtrsim 2.55$ \AA, the $0^+ \rightarrow 0^-$
frequency may not be realistically observable in a condensed phase because the
environment will decohere the system and suppress tunneling \cite{Weiss}.

For $R \approx 2.45$ \AA, there are some experimental data points that lie
between the two continuous theoretical curves.  
We consider three possible reasons for this discrepancy.
First, the one-dimensional potential may be unreliable in this
regime. However, we
consider this unlikely because the potential appears to successfully
describe so many other properties [bond lengths, geometric and frequency isotope effects]. Second, the two-dimensional character of the potential becomes
important [i.e., the coupling of X-H stretch with the X-Y stretch].
Third, there is significant uncertainty in the experimental values for
the frequency in this regime.
IR spectra for such strong H-bonded complexes  in this frequency
range are broad (compare Figure 2 in Reference \onlinecite{Novak})
and it is difficult to identify the
appropriate vibrational frequency \cite{Grdadolnik}. 
This large width is due to the combined effects of the large thermal and 
quantum fluctuations in $R$ (compare
Figure 6 in Ref. \onlinecite{Pirc2010}) and the
fact that the stretch frequency varies significantly with $R$.

The present results are relevant to infra-red spectra measured for ice under high
pressures, including the symmetric phase, Ice X \cite{goncharov,aoki}.  Two
vibrational modes are seen.  These can be identified with the curves for
$E_{1^+}-E_{0^-}$ and $E_{0^-}-E_{0^+}$ shown in Figure \ref{figfreq}.  Some caution
is in order in making a quantitative comparison because water does not have a
symmetric donor and acceptor for hydrogen bonding.  

For the rest of this manuscript, we refer to the $0^- \rightarrow 1^+$
transition frequency as the X-H stretch frequency, $\Omega$.

\section{Isotope effects}
\label{sec-isotope}

\subsection{Secondary geometric isotopic effects}

Figure \ref{figbondlength} and Section \ref{sec-bondlengths} discuss the primary geometric isotope
effects where the X-H bond length changes upon
substitution of the hydrogen with deuterium.
Secondary effects are those where the X-Y bond length changes, and are
also known as the Ubbel\"ohde effect \cite{Ubbelohde}.
 There have been extensive experimental
\cite{HamiltonAC63,LimbachJMS04,Limbach09,Ip12,Madsen07}
and theoretical 
\cite{LiPNAS11,Ichikawa,Saitoh,Tanaka,Sokolov1,Sokolov,Swalina,Vener07,Koval02,YangCPL,YangZPC} investigations of these geometric isotope effects. 

The secondary geometric isotope effect complicates the interpretation of other
isotope effects. Since the $R$ value changes between the isotopes, the effective
one-dimensional potential for each of them is different. Therefore, the shifts
due to the primary isotope effect are further modified. This convolution of
geometric isotope effects is seen by comparing the crystal structure of CrHO$_2$
and CrDO$_2$; in the former the O-H-O bond appears to be symmetric ($r=R/2$)
with $R = 2.49 \pm 0.02$ \AA, whereas the O-D-O bond is asymmetric with an O-D
bond length of $0.96 \pm 0.04$ \AA, with 
$R = 2.55 \pm 0.02$ \AA \cite{HamiltonAC63}. 
The shift of $R$ value appears small (a 2\% change) for the
low frequency motion that represents the X-Y stretch, but the effect is
palpable. 



Generally, one observes that for moderate H-bonds the equilibrium donor-acceptor
distance $R$ increases with substitution of hydrogen with deuterium
\cite{Ichikawa}. This is sometimes referred to as a positive secondary geometric
isotope effect.  For strong bonds, a negative effect, i.e., decrease of $R$, is
observed. For weak bonds, $R$ decreases, and understanding this requires
inclusion of the transverse vibrational modes \cite{LiPNAS11,Swalina}, as
discussed in Section \ref{sec-compete}.

We now consider a simple extension to our  model potential in
order to describe the secondary geometric isotope effect for strong to moderate
bonds. We draw from the model studies by Sokolov, Vener, and
Saval'ev \cite{Sokolov1}.  Consider a two-dimensional potential in terms of $r$
and $R$. This will contain an attractive (with respect to $R$) contribution from
the H-bond [$\epsilon_{-}(r,R)$ in our model] as well as a repulsive term
associated with the donor-acceptor repulsion \cite{repulsion} 
[so far not included in our model].  Competition between these two contributions
determines the \emph{classical} donor-acceptor bond length, here denoted $R_0$.
Such a two-dimensional potential would be the same for hydrogen and deuterium.
Here we may carry out a Born-Oppenheimer-like treatment of $r$ and $R$. Upon
taking an expectation value with the ground state vibrational wavefunction along
the fast coordinate, $r,$ an effective one-dimensional potential along $R$ is
obtained with the following form:
\begin{equation}
U_0(R) = U(R_0) + {K(R_0) \over 2} (R-R_0)^2 + Z(R),
\label{eqn-gse}
\end{equation}
where the first two terms on the right-hand side are a local quadratic expansion
about the $R_0$, and represent the elastic modulation of energy along the
donor-acceptor stretching coordinate. The essential physics of the isotope
effect is in the third term, the zero-point energy, 
\begin{equation}
Z(R) \equiv E_{0^+}(R) - \epsilon_{-}(r_{min},R),
\label{eqn-0pe}
\end{equation}
of the hydrogen (deuterium) motion. 

Note that $Z(R)$ is not required to be a minimum at $R_0$.  Minimising the total
energy (\ref{eqn-gse}) as a function of $R$ gives the equilibrium bond length 
\begin{equation}
R_{eq}= R_0 -  {1 \over  K(R_0)}  {d Z(R_0) \over d R}
\label{eqn-Req}
\end{equation}
to first order in $\hbar$ \cite{correction}. This equation was previously
presented by Sokolov, Vener, and Saval'ev \cite{Sokolov1}. They used
 zero-point energies obtained from different model potentials to that used here, and
they also assumed that $K(R)$ was constant. The physics involved is identical
to that used in solid state physics to calculate the effect of isotope
substitution on the lattice constant of a crystal \cite{Allen,Pamuk}.

We estimate $K(R_0)$, the elastic constant in the above model, from experimental
information in the article by Novak (Ref.~\onlinecite{Novak}, Figure 10 and
Table V).  It shows significant variation with $R_0$, increasing by a factor of
about 6 as $R_0$ decreases from 2.7 to 2.44 {\AA}.  The data fits an exponential
form,
\begin{equation}
K(R_0) = \bar{K} \exp\left[ -c(R_0-\bar{R}_0) \right], 
\label{eqn-Kfit}
\end{equation}
with
$\bar{K} = 
(55 \pm 3) \times 10^3 {\rm cm}^{-1}/\text{\AA}^2$, 
and $c=(7.3 \pm 0.8)$ \AA$^{-1}$, and $\bar{R}_0 \equiv 2.5$ \AA.

\begin{figure}[htb] 
\centering 
\includegraphics[width=82mm]{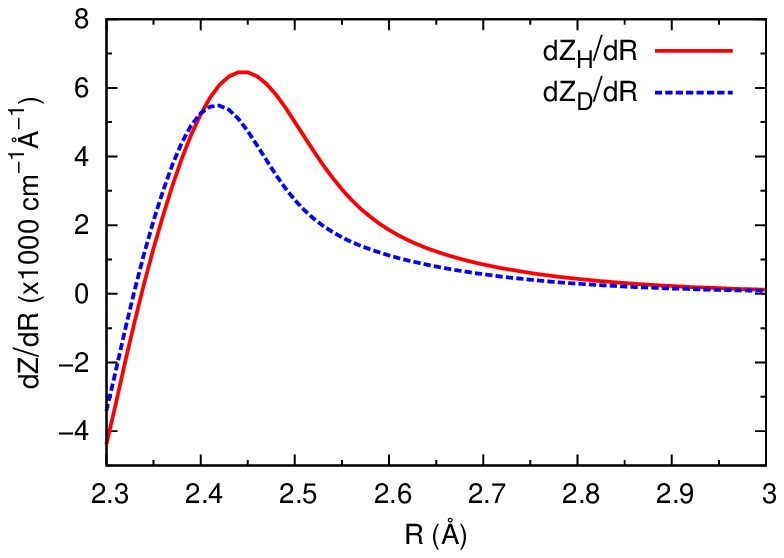}
\includegraphics[width=88mm]{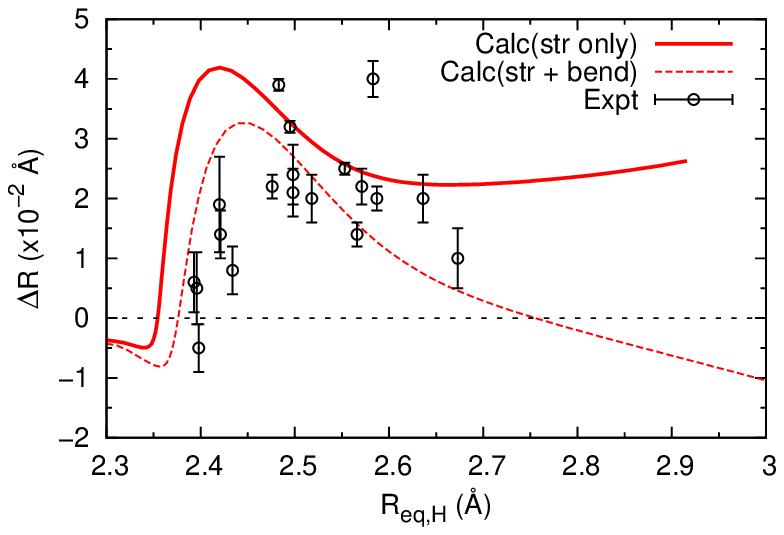}
\caption{
Non-monotonic dependence of the secondary geometric isotope effect on the
donor-acceptor distance.  The top panel shows the slope of the zero-point energy
in cm$^{-1}$/\AA.  The red curve is for hydrogen and the dashed blue curve for
deuterium.  Note that the maxima occur at different values of $R$ and that the
curves cross for $R \simeq 2.4$ \AA.  The difference between the two curves
determines the secondary geometric isotope effect [compare equation
(\ref{eqn-gie})] which is shown as the solid curve in the bottom panel. The
inclusion of the zero-point contribution of the bend modes, discussed in Section VII
and based on results in
Ref.~\onlinecite{McKenzieCPL}, yields the dashed curve in the bottom panel. Experimental data are taken
from Table 1  in Ref. \onlinecite{Sokolov1}.
}
\label{fig-gie}
\end{figure}

The zero-point energy of the X-H stretch is a non-monotonic function of
$R$.  The significant
variation with $R$ reflects the qualitative changes
in the one-dimensional potential that occur as one changes
from weak to moderate to strong bonds (compare Figure 3).
Furthermore, there are subtle differences between hydrogen and deuterium
isotopes. The top part of Figure \ref{fig-gie} shows a plot of the slope $dZ/dR$
versus $R$ for both hydrogen and deuterium. This slope is small and positive for
large $R$, increases as $R$ decreases until it reaches a maximum for $R \simeq
2.45$ {\AA} for hydrogen ($R \simeq 2.40$ {\AA} for deuterium), becomes zero for
$R \simeq 2.33$ \AA, and turns negative for smaller $R$.

With the above pieces of information, the secondary geometric isotope effect
is given by
\begin{equation}
\Delta R \equiv R_{eq,\,D} - R_{eq,\,H} =  
{1 \over  K(R)}\left( 
{d Z_H \over d R} 
-
{d Z_D \over d R} 
\right).
\label{eqn-gie}
\end{equation}
In this equation, the $0$ subscript for $R$ has been dropped.  $R_0$ was used
earlier in the section to indicate the classical minimum for various complexes.
However, $R$ in the model effectively takes the role of $R_0$, scanning through
the classical minima of all complexes. The solid curve in the bottom part of
Figure \ref{fig-gie} (labelled `str only') shows a plot of $\Delta R$ vs
$R_{eq,H}$, including a comparison with experimental data from a wide range of
complexes, as tabulated in Ref.~\onlinecite{Sokolov1}.

We point out that the
 $x$-axis of this plot, obtained from equation (\ref{eqn-Req}),
is different from $R$, the minimum of the classical potential.
 For $R \geq 2.35$ \AA, we find that $R-R_{eq,H}$ is,
like $Z(R)$, a non-monotonic function of $R$. It reaches
a maximum value of 0.1 {\AA} for
$R\simeq 2.5$ \AA, and drops towards zero on either side. Therefore, there is an
important difference between plotting the secondary geometric isotope effect
versus $R_{eq,H}$ and versus $R$. The former is the 
most self-consistent approach, since
$R_{eq,H}$ is what is experimentally measured.
But both approaches produce qualitatively similar results.

The proximity of the theoretical prediction by Eq.~\eqref{eqn-gie} to
experimental data is encouraging. In particular, the model predicts negative
$\Delta R$ values at short donor-acceptor distances (strong H-bonds), though
the location of the sign change is slightly offset from experimental data. The
fact that the secondary geometric isotope effect becomes negative (and small)
for strong H-bonds is seen in {\it ab initio} molecular dynamics simulations for
H$_5$O$_2^+$ \cite{Suzuki2013}, H$_7$N$_2^+$ \cite{YangCPL}, and H$_3$F$_2^+$
\cite{Suzuki2012}.  For example, for H$_5$O$_2^+$, it is found that at a
temperature of 100 K, $R = 2.417$ {\AA} and $\Delta R = -0.004$ {\AA}
\cite{Suzuki2013}.

The value of $R_{eq,H}$ from Eqn. (\ref{eqn-Req}), and consequently
 $\Delta R$, depends on the \emph{total} zero-point energy, i.e., not just that from the
X-H stretch but also from the X-H bends. It was recently
noted \cite{LiPNAS11} that the influence of the bends is rather pronounced at large X-Y
separations ($\simeq 2.7$ \AA). The dashed line labelled `str+bend' in the lower
panel of Figure \ref{fig-gie} is an attempt to include the effect of
the zero-point energy of the bending vibrations
as well, and is discussed in detail in Section \ref{sec-compete}.

\subsection{Vibrational frequency isotope effects}

The ratio of the frequency of longitudinal X-H stretching mode for hydrogen to
deuterium isotopes is observed experimentally to
be a non-monotonic function of the X-Y distance
with values varying  between 0.85 and 2.0 \cite{Novak,Sokolov,Grech,sponge}.
In contrast, for the torsional/bending modes, the isotope effects are trivial.
Table 6 of Ref. \onlinecite{Novak} shows that as the O$\cdots$O distance increases from 2.44 {\AA}
to 2.71 {\AA}, the ratio of the O-H to O-D (out of plane)
bend frequencies vary little, lying in
the range 1.32-1.44, and show no significant trend.  Broadly, they are
consistent with the semi-classical harmonic ratio $\sqrt{2}$.  This is
expected since for the bending mode there is no significant anharmonicity
(compared to the stretch mode).  With
respect to the $\phi$ co-ordinate in Figure 1, hydrogen bonding simply hardens
the potential for non-linear arrangements.

\begin{figure}[htb] 
\centering  
\includegraphics[width=80mm]{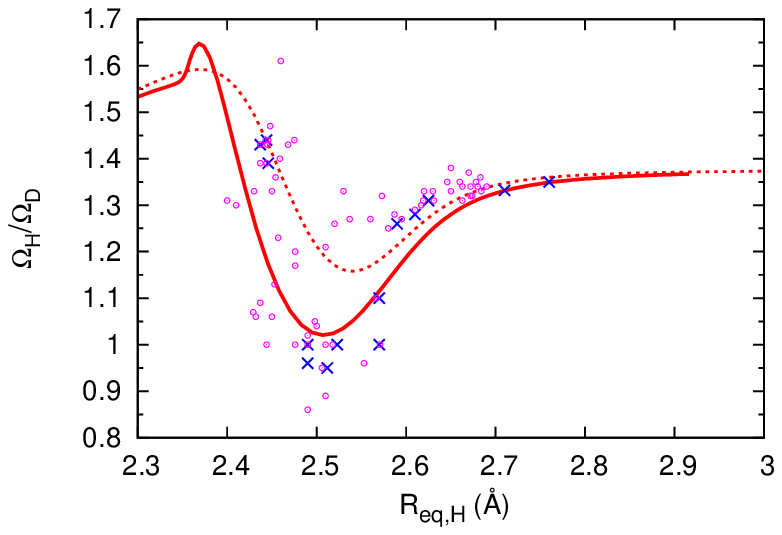}   
\includegraphics[width=80mm]{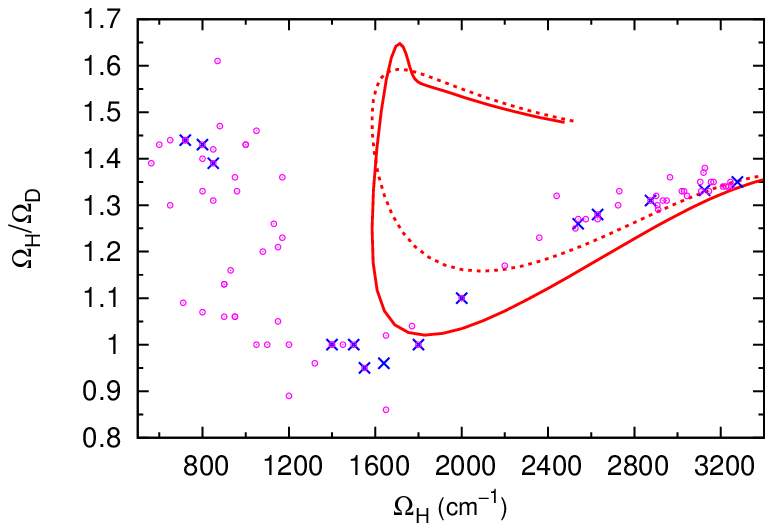}   
\caption{
Top panel: Correlation between the frequency isotope effect and the
donor-acceptor distance $R_H$, of the hydrogen isotope.  The vertical axis is
the ratio of the O-H stretch frequency to the O-D stretch in the same compound.
The solid curve is $\Omega_H(R_H)/\Omega_D(R_D)$ (i.e., the two frequencies are
calcuated for different one-dimensional potentials) whereas the dashed curve is
the frequency ratio calculated at the same distance $R_H$ (i.e., for the same
potential).  The difference between the two curves for $R \sim 2.5$ {\AA},
highlights the contribution of the secondary geometric isotopic effect,
calculated from equation (\ref{eqn-gie}). Bottom panel: The data is the same as
in the top panel, but the horizontal axis is $\Omega_H$ instead of $R_H$.
Experimental data in both plots are from Table 6 in Ref.
\onlinecite{Novak} (crosses) and Table 1  in Ref. \onlinecite{Sokolov} (open
circles).
}
\label{figisotope}
\end{figure}

Figure \ref{figisotope} compares the calculated correlation between the
frequency isotope effect with the donor-acceptor distance for the hydrogen
isotope for a wide range of complexes.  It is particularly striking that if one
simply calculates the frequencies for hydrogen and deuterium isotopes at the
{\it same} donor-acceptor distance [i.e., with the same one-dimensional
potential] one does not obtain quantitative agreement with the experimental data
for $R_{eq,H} \simeq 2.4-2.5$ {\AA} (compare the dashed curve in Figure
\ref{figisotope}).  Instead, one needs to take into account the secondary
geometric isotope effect and calculate $\Omega_D$ at $R=R_{eq,D}$, given by equation
(\ref{eqn-gie}) and plotted in Figure \ref{fig-gie} (lower panel, solid curve).
For $R_{eq,H} \simeq 2.4-2.5$ {\AA}, the secondary geometric isotope effect is
largest, $\Delta R \simeq 0.03$ {\AA}. Although this change  in value is small
relative to $R$, it makes sufficient alterations in the one-dimensional
potential along $r$ for deuterium so that $\Omega_D(R_{eq,D})$ becomes comparable to
$\Omega_H(R_{eq,H})$ (their ratio is closer to 1). Previously, Romanowski and Sobczyk
\cite{Romanowski} calculated a curve similar to the dashed one shown in Figure
\ref{figisotope} and suggested that the discrepancy with the experimental data
may be due to a change in the potential associated with the secondary geometric
isotope effect. We have shown that this is indeed the case.
We also note that the horizontal axis $R_{eq,H}$ is given by equation (9).

The $\Omega$ value for both isotopes are calculated as $E_{1^+}-E_{0^-}$
energies. At short O$\cdots$O distances $\lesssim 2.5$ \AA, the $E_{0^-}-E_{0^+}$
(ground state tunneling splitting) frequencies also enter the range of the
experimental data. However, the ratio of these frequencies for hydrogen and
deuterium lie above 1.5 (not shown), and thereby above the available
experimental data for O-H$\cdots$O.
However, experimental data for the frequency ratio of N-H$\cdots$N 
systems does increase up to 2 for short bonds \cite{Grech}.

An alternate way of examining the isotope effects with the same data is with a
plot of the ratio $\Omega_H/\Omega_D$ against $\Omega_H$
 rather than $R_{eq,H}$
\cite{Novak,Sokolov,Grech}.
 This is done  
in the lower panel of Figure \ref{figisotope}. The present model's predictions
without (dashed curve) and with (solid curve) secondary geometric isotope effect
corrections, deviate significantly from the experimental plot, 
with particularly
strong deviations for $R_{eq,H} \lesssim 2.5$ \AA, where the frequency ratio
turns upward for $\Omega_H \simeq 1700 \ {\rm cm}^{-1}$ . Note, however, that continuous
curves do capture the range of the frequency ratios, just as they do in the
upper panel of the figure. The discrepancy is due in part to $\Omega_H$; it does not
take on values as low as reported in experiments for $R \lesssim 2.5$ \AA, an
observation noted previously for Fig. \ref{figfreq}. 
In effect, the H-bond potential model in
this work is able to recover frequency ratios rather well, but not the
experimental frequencies in certain strong H-bonding regions. Then again, the
experimental frequencies in this range are difficult to unambiguously identify;
see Section \ref{sec-freq} above as well as Refs.
\onlinecite{Grdadolnik} and
\onlinecite{Pirc2010}. 



\subsection{Simple models for frequency isotope effects}
\label{sec-simple}

Some insight can be gained into the variation of the isotope
effects with the donor-acceptor distance by considering
analytical results for simple model potentials that are
relevant in different limits.

\begin{enumerate}
\item \emph{Harmonic potential}\\
This approximately describes weak hydrogen bonds.  For a potential
$V(r)=A(r-r_0)^2/2$, the energy eigenvalues are $E_n=(n+1/2)\hbar \sqrt{A/M}$.
Hence, the frequency $\Omega \equiv (E_1-E_0)/\hbar$.  The ratio of the
frequencies for hydrogen and deuterium is
\begin{equation}
{\Omega_H \over \Omega_D} = \sqrt{2}=1.41.
\end{equation}
For weak bonds the anharmonicity factor is small enough ($\chi \sim 0.03$ in equation
\ref{morse_freqratio} below) that the above ratio is a reasonable approximation.

\item \emph{Morse potential}\\
This approximately describes the anharmonicity associated with (parameterized
around) the bottom of the
potential for moderate to weak hydrogen bonds.  This is not to be confused with
the Morse potential that we use to describe the diabatic states.
For a Morse
potential, the eigenvalues are $E_n = (n+\tfrac{1}{2}) \hbar\omega_0 -
(n+\tfrac{1}{2})^2 \hbar\omega_0 \chi,$ where 
$\omega_0$ is the harmonic frequency 
and $\chi \equiv \hbar\omega_0/4D_0$ is the
anharmonicity. The transition frequency is then $\Omega = (E_1-E_0)/\hbar =
\omega_0(1-2\chi)$.
The ratio of the hydrogen and deuterium frequencies is
\begin{equation}
\label{morse_freqratio}
{\Omega_H \over \Omega_D} = \sqrt{2}
\left( {1 - 2\chi} \over {1 - {\sqrt{2}\chi} }\right).
\end{equation} 
Hence, as $R$ decreases, we expect the frequency ratio to decrease as is
observed.  Even for large anharmonicity ($\chi \sim 0.2-0.25$), the ratio only
decreases to about 1.1-1.2, as is observed in the full calculation (compare
the dashed curve in the upper panel of
Figure \ref{figisotope}).

\item \emph{Infinite square well potential}\\
For strong bonds, the potential is approximately a square well of width
$L=R-2r_0$.  This observation was pointed out in References 
\onlinecite{Horbatenko} and
\onlinecite{Belot}.
For a well of width $L$, the energy of the $n$-th
level is 
\begin{equation}
E_n =   {\hbar^2 n^2 \over 2 M L^2 }
\end{equation} 
where $n=1,2,3,\cdots$.
The ratio of the frequencies for the two isotopes is then
\begin{equation}
{\Omega_H \over \Omega_D} = 2.
\end{equation} 

\end{enumerate}

The detailed calculations of the isotope frequency ratio shown in Figure
\ref{figisotope} are consistent with the above three limits.  As $R$ decreases,
$ \Omega_H / \Omega_D$ decreases below 1.4 reaches a minimum, and then for $R$
values corresponding to
a single well potential the ratio increases to values larger than $1.4$.

The ground state wave function for the infinite square
well is ($n=1$) 
\begin{equation}
\Psi_g(r) = \sqrt{2 \over L}\sin\left({\pi r \over L} \right),
\end{equation} 
which is independent of the mass $M$. Hence, the zero-point
energy depends on $1/M$, but the zero-point motion is {\it
independent of $M$}. This is in contrast to the case of a  harmonic potential
for which the zero-point wavefunction does have $M$
dependence.
Indeed, this explains why calculations of the ground state probability
distribution $|\Psi_g(r)|^2$ for the protonated ammonia dimer N$_2$H$_7^+$
\cite{YangCPL}, for H$_3$O$_2^-$ \cite{YangZPC}, and for sodium bihydrogen
bisulfate \cite{Pirc2010} found virtually identical probability distributions
for both isotopomers.

The non-monotonic dependence of the zero-point energy $Z(R)$ on $R$ 
(compare the upper panel of Figure \ref{fig-gie}) can also be
understood in terms of the analytic limits discussed above.  As $R$ decreases
the potential gets more anharmonic and the zero-point energy
$Z=E_0=\tfrac{1}{2}\hbar \omega_0(1-\hbar\omega_0/8D_0)$ decreases because the
effective $D_0$ of the local Morse potential
also decreases.  However, in the single well regime, $Z(R) \sim
1/(R-2r_0)^2$, and the zero-point energy increases with decreasing $R$.

\section{Competing quantum effects}
\label{sec-compete}

There is a torsional or bending vibration associated with periodic oscillation
of the angle $\phi$ shown in Figure \ref{fig1}.  This is related to the
libration mode in water and ice.  The bending vibrations make an important
contribution to the total zero-point energy of the system [compare equations
(\ref{eqn-gse}) and (\ref{eqn-zpe}) below].  As the donor-acceptor distance
decreases, the bending frequency and the associated zero-point energy increase
(compare Figure 6 in Ref.~\onlinecite{McKenzieCPL}). This is the opposite trend
to the X-H stretch. These opposite dependences lead to the notion of competing
quantum effects \cite{Habershon,Markland,Romanelli}.

In this section, we make a preliminary analysis of the role of H-bond bending on
the secondary geometric isotope effects.  The total zero-point energy is
\begin{equation} Z(R) \equiv
Z_\parallel(R) + Z_{\perp,o}(R)+ Z_{\perp,i}(R). \label{eqn-zpe} \end{equation}
The terms are the zero-point energy associated with X-H vibrations parallel to
the hydrogen bond (stretch), out-of-plane bend ($o$), and in-plane
bend ($i$) of X-H$\cdots$Y. 
In the diabatic state model \cite{McKenzieCPL}, 
the effect of H-bonding on hardening of the two bend motions is similar,  
\begin{equation} 
\Omega_{\perp,o/i}(R)^2 
=
\omega_{\perp,o/i}^2 + 2 f(R)
\label{eqn-bend} 
\end{equation}
where 
$\omega_{\perp,o/i}$ is the frequency in the absence
of an H-bond and
the function $f(R)$ is given in Eqn. (6) of 
Ref.~\onlinecite{McKenzieCPL}. At least in the $R$ range of
interest, $f(R)$ is a positive function that monotonically decreases with
increasing $R$: $f'(R)<0$.
In general $ \omega_{\perp,i} > \omega_{\perp,o}$ and so
$ \Omega_{\perp,i} > \Omega_{\perp,o}$.        
The  contributions of the two bending bending modes
to the zero-point energy (\ref{eqn-zpe}) are taken to be
$Z_{\perp,o/i}=\tfrac{1}{2}\hbar\Omega_{\perp,o/i}(R)$: they are treated as
harmonic oscillators.

The frequency    $\Omega_{\perp,o}(R)$ is taken from
(the solid line of) Figure 6 of Ref.~\onlinecite{McKenzieCPL}.
Little data is available for the in-plane bend
$\Omega_{\perp,i}(R)$
because the interpretation of experimental data is difficult due to
the strong mixing of this mode with others \cite{Novak}. 
Hence, we use the following simple analysis to estimate its effect.
If we take the derivative of Eqn. (\ref{eqn-bend}) with respect to $R$ we
obtain,
\begin{equation} 
\Omega_{\perp,o} 
\frac{d\Omega_{\perp,o}}{dR}
=
\Omega_{\perp,i} 
\frac{d\Omega_{\perp,i}}{dR}
=
{df \over dR}.
\label{eqn-bend2} 
\end{equation}
Hence, we can write the derivative of the total bend ZPE
\begin{equation} 
\frac{dZ_{\perp}}{dR}
=
\frac{d Z_{\perp,o}}{dR}\left(1 + 
\frac{\Omega_{\perp,o}}{\Omega_{\perp,i}}
\right).
\label{eqn-bend3} 
\end{equation}
It can be seen from equation (\ref{eqn-bend}) that
$1 > \Omega_{\perp,o}(R)/\Omega_{\perp,i}(R) > \omega_{\perp,o}/\omega_{\perp,i} $, and that this
frequency ratio 
progressively increases towards unity as $R$ decreases. Given that information about the
out-of-plane bend is known better than the in-plane bend, we make a limiting approximation that 
$\Omega_{\perp,o} \simeq \Omega_{\perp,i},$ so we can write
\begin{equation}
\frac{dZ_{\perp}}{dR}
\simeq 2
\frac{d Z_{\perp,o}}{dR}.
\end{equation} 
This becomes less reliable for larger
$R$ (when $f(R)$ becomes smaller), giving an overestimate of the
magnitude of the total bend derivative.

All terms in equation \eqref{eqn-zpe} vary significantly with $R$ in
the range of interest (2.3-3.0 \AA). The first term has a
non-monotonic trend (as shown in the upper panel of
Figure \ref{fig-gie}), whereas the bend terms
decrease monotonically as $R$ is increased. So the total zero-point energy
involves a subtle competition between the stretch and bend components at different values
of $R$.

The net secondary geometric isotope effect comes from a balance between 
$\left[dZ_{\parallel,H}/dR - dZ_{\parallel,D}/dR\right]$ for the stretch and
$2\left[dZ_{\perp,o,H}/dR - dZ_{\perp,o,D}/dR\right]$ for both bends together
(compare equation \eqref{eqn-gie}).  Noting that $\Omega_{\perp,o/i}$ scale
essentially as the square root of the mass of H or D, the derivative difference for
the bends can be simplified to 
$2\left(1-\tfrac{1}{\sqrt{2}}\right) dZ_{\perp,o,H}/dR$.

At $R \simeq 2.4$ {\AA} and $R \gtrsim 2.7$ \AA, the derivative difference for
the stretch mode is small; see the upper panel of Figure \ref{fig-gie}. It is in
these regions that the the bend contributions will be particularly noticeable.
For example, at 2.4 \AA, $dZ_{\perp,o,H}/dR \simeq -800$ cm$^{-1}$, so that the
derivative difference for both bends together is about $-450$ cm$^{-1}$. The
secondary geometric isotope effect is negative in sign at this $R$ value, and contains
a substantial contribution from the bend.

The dashed line in the lower panel of Figure \ref{fig-gie} (labelled `str+bend')
gives an estimate of the secondary geometric isotope effect including
both bends. The overall features of the change in donor-acceptor
distance $\Delta R$ are not too different at short
$R$, apart from an overall downward shift.  But at $R_{eq,H} \gtrsim 2.7$ \AA, the bend
contribution overtakes the stretch giving rise to a negative $\Delta R$. The
position of the crossover may change slightly with a more refined treatment of
the bend modes and the model itself. 
Specifically, for weak bonds the contribution of the in-plane bend will
become smaller than that of the out-of-plane bend 
(compare Eqn. (\ref{eqn-bend3})).
Indeed, this difference was also found for path integral
simulations of isotopic fractionation in water \cite{Markland}.
Consequently, $\Delta R$ will become negative at a larger $R_{eq,H}$ than
the value of about 2.7 \AA \ shown in our Figure 7.
At still larger distances,
$R_{eq,H} > 3.0$ \AA, the H-bonding becomes very weak, and it is expected that the
$\Delta R$ curve eventually goes to zero.

The qualitative aspect of a negative
$\Delta R$ for weak H-bonds is in agreement with recent work 
of Li {\it et al.} \cite{LiPNAS11} 
based on Path Integral Molecular Dynamics simulations. They
showed that the bend modes would dominate over the stretch
for weak H-bonds, leading to a negative secondary geometric isotope effect at
large donor-acceptor distances. They found a change in the sign of the geometric isotope effect
when the H-bond strength was such that the X-H stretch frequency
was reduced by about 30 per cent. From Figure \ref{figfreq} we estimate
this corresponds to  $R _{eq,H} \simeq 2.6$ \AA, a value somewhat lower
than the crossover region we see in Figure \ref{fig-gie}.

\section{Possible future directions}

There are several natural directions to pursue 
future work. 
These include the description of asymmetric complexes
where the proton affinity of the donor and acceptor are
different. As a result the one-dimensional potential
is no longer symmetric about $r=R/2$.
Development of a full two-dimensional potential $V(r,R)$
will allow treatment of the secondary geometric isotope effect
without introducing the empirical elastic constant $K(R)$ and
investigating of the coupling of thermal and quantum fluctuations
between $R$ and the X-H stretch.
This simple diabatic state model approach can be readily be applied
to more complex H-bonded systems such as those
associated with
solvated Zundel cations \cite{Reed},
 excited state proton transfer,
double proton transfer in porphycenes \cite{Waluk}, and
water wires \cite{Chen}.
Finally, we briefly discuss two other future directions.

\subsection{Anisotropic Debye-Waller factors}

For crystal structures, one assigns ellipsoids
associated with the uncertainty of the positions of
individual atoms determined from X-ray or neutron diffraction
experiments. The relevant quantities are known
as Atomic Displacement Parameters or Debye-Waller factors. In the 
absence of disorder, their
magnitude is determined by the quantum and thermal fluctuations
in the atomic positions. 
Anisotropy in the ellipsoid reflects a directional dependence
of bonding and the associated vibrational frequencies.
Anisotropy in the associated kinetic energy of protons in liquid
water and in ice was recently measured by inelastic neutron scattering
\cite{Romanelli}.

The variation in the anisotropy of the ellipsoid with donor-acceptor distance
has been calculated for ice by Benoit and Marx \cite{Benoit05}.  Anisotropy of
the Debye-Waller factor for the position of protons in enzymes has recently been
critically examined  with a view to identifying low-barrier H-bonds
\cite{Hosur}.  The authors found that anisotropy is correlated with the presence
of short bonds and with ``matching $pK_a$'s'' [i.e., the donor and acceptor have
similar chemical identity and proton affinity], as one would expect.

Our calculations of the variation of X-H stretch zero-point energy with 
respect to $R$ and the X-H bend frequency 
(Figure 6 in Ref. \onlinecite{McKenzieCPL}) suggest
the anisotropy has a non-monotonic dependence on $R$.

\subsection{Hamiltonian for non-adiabatic effects}

The model Hamiltonian (\ref{eqn-ham}) has a natural
extension to describe non-adiabatic effects associated
with a quantum mechanical treatment of 
the hydrogen atom co-ordinate $r$.  
The {\it harmonic limit} for symmetric donor
and acceptor 
 corresponds to a 
spin-boson model \cite{KoppelACP84} with the quantum Hamiltonian
\begin{equation}
H = { \hat{p}^2 \over 2M} + {M \over 2} \omega^2 \hat{q}^2
+
\left(\begin{array}{cc} 
g \sqrt{2M\omega} \hat{q}          & 
\Delta(R) \\
\Delta(R)  & 
 - g  \sqrt{2M\omega}\hat{q}
\end{array}\right)
\label{eqn-hamna1}
\end{equation}
where $\hat{p}$ is the momentum operator, conjugate to $\hat{q} \equiv \hat{r}-R/2$, and $ g \equiv \sqrt{M \omega^3 \over 2}
(R/2-r_0)$.
This Hamiltonian can be re-written as
\begin{equation}
H = \Delta  \sigma_x  
+ g (a^\dagger + a) \sigma_z
+ \omega a^\dagger a
\label{eqn-hamna}
\end{equation}
where $\sigma_x$ and $\sigma_z$ are Pauli matrices
and $a$ and $a^\dagger$ are annihilation and creation
operators, respectively, associated with the $r$ co-ordinate.
This Hamiltonian has an
analytical solution in terms of continued fractions \cite{PaganelliJPCM06}.

The fully quantum Morse potential has an exact analytical solution and 
an algebraic representation in terms of creation and annihilation
operators \cite{Levine}.  
Hence, an algebraic treatment of the quantum version of the model Hamiltonian
(\ref{eqn-ham}) (i.e. without taking the harmonic limit)
may also be possible, because the off-diagonal terms are
independent of $r$. Given the quantitative importance of the
anharmonicity associated with the Morse potential \cite{note} this is desirable.

Previous studies \cite{McKemmish,Stanton}
of the Hamiltonian (\ref{eqn-hamna})
suggest that the most significant deviations from
the Born-Oppenheimer approximation will occur when 
the bare vibrational frequency $\omega \sim \Delta(R)$ and 
also the barrier height.
This will occur when $R \sim  2.5$ {\AA}.

\section{Conclusions}

We have clearly shown that the quantum motion of the proton has a significant
effect on the properties of H-bonds of strong to moderate strength between
symmetric donor (X) and acceptor (Y) groups.  A simple one-dimensional potential
for the linear transfer path (X-H stretch) of the proton at various
donor-acceptor separations ($R$), based on a two-diabatic state model with only
a very few parameters, was used for this purpose. The structure of this
potential varies from a high-barrier double-well for weak and moderate H-bonds
($R\gtrsim2.7$ \AA) to a single well for strong H-bonds ($R \sim 2.4$ \AA).  Our
analysis of the proton motion on this potential gives qualitative 
and quantitative descriptions of several
correlations as a function of $R$ for O-H$\cdots$O containing materials.

The model's predictions of the basic properties of hydrogen bonding, viz.~X-H
bond length variations and vibrational frequency red-shifts, for both hydrogen
and deuterium isotopes, compare well with known experimental information over a
wide $R$ range. The key additional prediction using a slight extension of the
model is that of the secondary geometric isotope, or Ubbel\"ohde, effect,
wherein the donor-acceptor distance is changed due to H to D isotopic
substitution.

The Ubbel\"ohde effect is a quantum effect whose magnitude depends on zero-point
energies (ZPEs) in the proton and deuteron's degrees of freedom. We have shown that the ZPE
along the X-H(D) stretch is able to capture the experimental trends for strong
H-bonds. The model potential shows qualitative changes for 
$R \lesssim 2.5$ \AA, when the barrier becomes comparable to or lower than the ZPE
of the X-H(D) stretch mode. Concomitantly, significant variations in the
\emph{difference} between the X-H and X-D  ZPE derivatives with $R$ are
observed in our model, which dominates the secondary
geometric isotope effect for strong to moderate H-bonds. This effect modulates
and, indeed, improves the model's predictions of the primary geometric isotope
effect as well.

In this paper, we have employed mainly one-dimensional quantum calculations
along the X-H stretch with the donor-acceptor distance ($R$) as a control parameter.
This alone is found to be quite insightful. Of course, higher dimensional
quantum treatments that include the X-H bends and $R$ are anticipated to yield
a still better quantitative description. Taking a short step in this direction, we have made a
preliminary analysis of the effect of the X-H bends in the context of the
Ubbel\"ohde effect. We find, in agreement with other recent works, that their
influence is mainly in the moderate to weak  H-bond regime, where
they begin to overtake the influence of the X-H stretch.

%
%

\begin{acknowledgments}

We thank E. Arunan, 
T. Frankcombe,
J. Grdadolnik, 
A. Hassanali, J. Klinman,
T. Markland, A. Michaelides, J. Morrone, J. Stare, and L. Wang
 for helpful discussions.
M. Ceriotti, D. Manolopoulos, T. Markland, and S. McConnell provided helpful
comments on a draft manuscript.
RHM received financial support from an
  Australian Research Council Discovery Project grant (DP0877875)
and an Australia-India Senior Visiting Fellowship.

\end{acknowledgments}


\begin{thebibliography}{39}
\bibitem{reduced}
A simple choice of reduced mass would be that of OH or OD diatoms. However,
since a double-well structure is involved, the mass of the proton motion is
taken to be the mass of the asymmetric stretch coordinate in an
O$\cdots$H$\cdots$O triatomic system: $1 / M = 1/m_H + 1/4\mu_{OO}$ and likewise
for the deuterium, where $\mu_{OO}=\tfrac{1}{2}m_O$. The numerical difference
between this choice [$M=32/33=0.967$ for protons] 
and the diatom variant is under 5\% for both H and D, so the
trends presented in this paper are not changed by either choice.
\bibitem{note}
It can be shown that the condition for no energy barrier
in the ground state is
\begin{equation}
\frac{\Delta(R)}{D}
 \geq \frac{(1-\exp(-x))^2}{1-\frac{1}{2}\exp(x)}
\end{equation}
where $x \equiv a(R/2-r_0)$.                
This leads to a symmetric X-H-Y bond in the adiabatic
limit when  the nuclear degrees of freedom are treated classically. 
Quantum chemistry calculations \cite{Limbach09} suggest
that for O-H-O complexes symmetric bonds occur when $R \simeq 2.4$ {\AA}.
Using this value and $a \simeq 2.2 \text{\AA}, r_0=0.96$ {\AA} gives
 $x = 0.52$,
and the above expression gives $\Delta(R) \simeq D$.
As an aside, we note that in the harmonic limit (i.e., $ x \ll 1$)
 the right hand side of  the above expression 
reduces to $ 2x^2 $ (harmonic limit) which for
$x = 0.52$ gives $\Delta \simeq 0.6D$.
 This  significant difference
from the value above      shows   
 the importance of using a full Morse potential for the diabatic
state energies.
\bibitem{harmonic}
Note that the harmonic curve shown in Figure 6 disagrees with
the curve shown in Ref. \onlinecite{McKenzieCPL}. Unfortunately, the
computer code used to calculate the latter contained an error.
\bibitem{repulsion}
A simple parametrisation of this will involve the steric term associated with
the Pauli repulsion between the electron clouds of the donor
and acceptor atoms \cite{Weinhold} which decays exponentially
with $R$. One also needs to include the effects of electrostatic interactions,
that become increasingly important at larger $R$ \cite{SagnellaJCP98}.
\bibitem{sponge}
A study \cite{Bienko} of the proton sponge, 
2,7-dibromo-1,8-bis-dimethylamino-naphthalene (Br$_2$DMAN),
found no secondary geometric isotope effect
[the N$\cdots$N distance was $R= 2.547$ {\AA}],
and intense IR modes at 589 and 284 cm$^{-1}$ for
H and D, respectively.
This gives an isotope frequency ratio of 1.67.
These experimental results are consistent with the potential
that we use for O-H$\cdots$O bonds with $R=2.4$ {\AA}.
A similar correspondence was pointed out in
Ref. \onlinecite{Grech}.
\bibitem{correction}
The validity of (\ref{eqn-Req}) requires that
$K(R) \gg {d^2 Z \over d R^2}(R)$. 
Estimating the curvature from Figure \ref{fig-gie}
we see the two terms are of comparable magnitude for $R \simeq 2.4, 2.55$ {\AA}.
However, we have parametrised $K(R)$ from experimental data
for X-Y stretch frequencies, not classical calculations, which
implicitly includes the second order corrections, i.e.,
$K_{eff}(R) = K(R) +  {d^2 Z \over d R^2}(R)$. 
But this highlights how the X-Y stretch frequency for moderate to strong
bonds may have a significant contribution from the zero-point
energy of the X-H stretch and exhibit a surprisingly large
dependence on hydrogen-deuterium substitution.
\end{thebibliography}
\end{document}